\documentclass[%
preprintnumbers,
nofootinbib,
 amsmath,amssymb,
 aps,
]{revtex4-2}

\usepackage{graphicx,color}

\usepackage{dcolumn}
\usepackage{bm}
\usepackage{mathtools}
\usepackage{here}

\begin{document}

\preprint{KEK-TH-2728 and J-PARK-TH-0317}

\title{Unveiling the role of vector potential in the Aharonov-Bohm effect}



\author{Masashi Wakamatsu}
\email[]{wakamatu@post.kek.jp}
\affiliation{KEK Theory Center, Institute of Particle and Nuclear Studies, High Energy Accelerator
Research Organization (KEK), 1-1, Oho, Tsukuba, Ibaraki 305-0801, Japan}



\begin{abstract}
The most popular interpretation of the Aharonov-Bohm (AB) effect is that
the electromagnetic potential locally affects the complex phase
of a charged particle's wave function in the magnetic field free region. 
However, since the vector potential is a gauge-variant quantity, 
not a few researchers suspect that it is just a convenient 
tool for calculating the force field. This motivates them to explain the AB effect
without using the vector potential, which inevitably leads to
some sort of non-locality. This frustrating situation 
is shortly summarized by the statement of Aharonov et al. that the AB effect may 
be due to a local gauge potential or due to non-local gauge-invariant fields. 
In the present paper, we shall give several convincing arguments, which 
support the viewpoint that the vector potential is not just a convenient 
mathematical tool with little physical entity. Despite its gauge arbitrariness, 
the vector potential certainly contains a gauge-invariant piece, which solely
explains the observed AB phase shift. Importantly, this component has a property
such that it is basically unique and cannot be eliminated by any regular gauge 
transformations. To make the discussion complete, we also discuss the role 
of remaining gauge arbitrariness still contained in the entire vector potential.
\end{abstract}


\maketitle

%

\section{Introduction}\label{sect1}

Since its first theoretical prediction \cite{ES1949},\cite{AB1959}, 
the Aharonov-Bohm effect has continued to provoke innumerable 
debates which concern the deep foundation of modern physics. 
(For review, see \cite{Peshkin1981}\nocite{OP1985}\nocite{PT1989}-\cite{WKZZ2018},
for example.)
A central question is whether the electromagnetic potential is a physical 
entity or it is just a convenient mathematical tool 
\cite{Feynman1963}\nocite{Konopinski1978}-\cite{Semon-Taylor1996}.
Without doubt, most popular interpretation of the AB-effect is that the magnetic
vector potential locally affects the complex phase of quantum electron wave 
function, thereby causing a change of phase that can be verified through 
interference experiments \cite{Tonomura1986},\cite{Osakabe1986}. 
Nonetheless, it is also true that there are many researchers who are not
completely satisfied with the vector potential explanation of the 
AB effect \cite{Healey2022}\nocite{AV2000}\nocite{ACR2016}-\cite{Heras-Heras2022}.
This is because the vector potential is a gauge-variant quantity with
inherent arbitrariness. By this reason, they believe that the vector 
potential is just a convenient mathematical tool for calculating the 
electromagnetic force field as is in fact the case with the classical electromagnetism. 
This motivate them to look for explanations which do not use the 
gauge-dependent vector potential \cite{Healey2022}\nocite{AV2000}-\cite{ACR2016}. 
One of the most influential studies along this line would be the work by
Vaidman \cite{Vaidman2012}, \cite{Vaidman2015}. 
He proposed an explanation for the AB effect via force between
the solenoid current and the moving electron rather than via electromagnetic
potential. Later, Vaidman's paper was criticized by 
Aharonov, Cohen and Rohrlich \cite{ACR2015C}. 
Through six thought experiments, these authors concluded that
the attempt to dispense with scalar and vector potentials is at least 
incompatible with the attempt to interpret the AB effect as a local effect. 
They also pointed out a potential problem inherent in the local force
explanation of the AB effect.
According to this force interpretation, the change in the phase of the electron's 
wave function implies a change in the electron's velocity, which appears to 
be incompatible with the so-called dispersionless nature of the 
AB effect \cite{Zellinger1986},\cite{Peshkin1999} \footnote{The dispersionless 
nature of the AB effect means that the observed AB phase shift is independent
of the velocity of the electron}.  
  
Partially motivated by Vaidman's work, several authors focused on
the interaction energy between the solenoid current and the moving
electron rather than the force between them \cite{Santos-Gonzalo1999}
\nocite{Kang2017}\nocite{Marletto-Vedral2020}\nocite{Saldanha2021}
-\cite{LHK2022}. (See also a related older work by Boyer \cite{Boyer1971}.)
They postulates that the change of the phase of the electron wave
function along a path is proportional to the change of the above
interaction energy along the same path.  
Then, by explicitly evaluating the change of the interaction energy
along a closed path of the electron, they showed that it reproduce
the standard answer for the AB phase shift.

Highly nontrivial is the validity or invalidity of these authors' further claim
as follows. According to them, since the change of the above interaction 
energy along the path of the electron is a gauge-invariant quantity, 
the {\it partial} AB phase shift along a non-closed is also a gauge-invariant 
quantity, which in turn implies that it can in principle be observed. 
However, in a recent paper \cite{Waka2024}, based on the framework
of a self-contained quantum mechanical treatment of the combined
system of a solenoid, an electron, and the quantized electromagnetic field,
we showed that there is no evidence to support their central claim,
i.e. the proportionality assumption between the interaction energy of the
solenoid current and the moving electron and the corresponding partial
AB phase shift. To summarize the situation to date, we feel that 
any attempt to dispense with the vector potential in the explanation of 
the AB effect has not reached a complete success. 
On the other hand, there already exist several works, which support 
physically reasonable nature of the vector potential interpretation \cite{Adachi1992}
\nocite{Stewart2003}-\cite{Li2012}. 
Only worry of this theoretical approach is the fact that
the vector potential is a gauge-dependent quantity, which makes it 
difficult to completely dispel some researchers' suspicion on its physical reality.
The purpose of the present paper is to get rid of these doubts
in the most convincing manner.

The remainder of the paper is structured as follows.
First, in sect.\ref{sect2}, we analyze the nature of the vector potential 
generated by an infinitely-long solenoid with the special intention of 
confirming its physically substantial nature.
It is demonstrated that the vector potential generated by an 
infinitely-long solenoid can be uniquely decomposed into the transverse
component and the longitudinal component, provided that 
physically unacceptable multi-valued gauge transformation is excluded.
To help understanding of the significance of the discussion in sect.\ref{sect3},
we provide in sect.\ref{sect4} a brief introduction to some past works which
tried to explain the AB effect, without using the standard vector potential
interpretation.
Also discussed there are the highly nontrivial claim in several recent 
literature, in which the authors insist that the partial AB phase-shift 
corresponding to a non-closed path of the electron can in principle
be observed, thereby proposing several concrete 
settings of measurement for verifying their 
claim \cite{Kang2017}\nocite{Marletto-Vedral2020}
\nocite{Saldanha2021}-\cite{LHK2022}.
Sect.\ref{sect5} summarizes the essential points of our vector potential 
interpretation of the AB effect improved upon in the present paper.

\section{On the vector potential generated by an infinitely-long solenoid}\label{sect2}

In order to discuss the essence of the AB effect in the simplest possible form,
it is customary to consider an idealized setting of infinitely-long 
solenoid with radius $R$ directed to the $z$-direction.
The stationary and uniform surface current distribution of the solenoid 
is represented as 
\begin{equation}
 \bm{J}_{ext} (\bm{x}) \ = \ B \,\delta (\rho - R) \,\bm{e}_\phi .
\end{equation}
Here we use the cylindrical coordinates $\bm{x} = (\rho, \phi , z)$ with
$\rho = \sqrt{x^2 + y^2}$ and $\phi = \arctan \left( \frac{y}{x} \right)$.
Our aim is to find the vector potential generated by the above
solenoid current distribution. 
There are various routes to reach this goal, but probably the most
instructive way is to start with the familiar Biot-Savart law 
represented as (note that we are basically handling magnetostatics)
\begin{equation}
 \bm{B} (\bm{x}) \ = \ \frac{1}{4 \,\pi} \,\nabla \times \int 
 \frac{\bm{j}_{ext} (\bm{x}^\prime)}{\vert \bm{x} - \bm{x}^\prime \vert} \,d^3 x^\prime .
\end{equation}
(For simplicity's sake, we use the Heaviside-Lorentz unit combined with the 
natural unit $\hbar = c = 1$.)
The importance of this formula is that the physical quantities contained in it 
(those are the external current distribution $\bm{j}_{ext} (\bm{x})$ and the generated 
magnetic field $\bm{B} (\bm{x})$) are all gauge invariant quantities. 
The form of the Biot-Savart law naturally promotes us to 
introduce the vector field $\bm{A} (\bm{x})$ by the relation
\begin{equation}
 \bm{B} (\bm{x}) \ = \ \nabla \times \bm{A} (\bm{x}).
\end{equation}
This quantity $\bm{A} (\bm{x})$ is nothing but what we call the (magnetic) vector 
potential. Naturally, the vector potential introduced in the above way is 
not unique. Its general form is given as
\begin{equation}
 \bm{A} (\bm{x}) = \bm{A}^{(S)} (\bm{x}) \ + \ \nabla \chi (\bm{x}), 
 \label{Eq:A_S+nabla_chi}
\end{equation}
with the definition
\begin{equation}
 \bm{A}^{(S)} (\bm{x}) \ \equiv \ \frac{1}{4 \,\pi} \,\int 
 \frac{\bm{j}_{ext} (\bm{x}^\prime)}{\vert \bm{x} - \bm{x}^\prime \vert} \,d^3 x^\prime ,
 \label{Eq:A_S_int}
\end{equation}
while $\chi (x)$ is an {\it arbitrary} scalar function \cite{Shadowitz1975}. 
The superscript $(S)$ on
$\bm{A}^{(S)}$ designates that it is a part of the vector potential $\bm{A} (\bm{x})$ which is
uniquely determined by the solenoid current distribution $\bm{j}_{ext} (\bm{x})$, 
provided that the relevant spatial integral in (\ref{Eq:A_S_int}) 
converges \footnote{Naively, this integral diverges, but it is known to converge
by using an appropriate limiting procedure \cite{Shadowitz1975}}. 
Arbitrary nature of the part $\nabla \chi (\bm{x})$ is interpreted as gauge degrees
of freedom of the vector potential.
This is of course a well-known story, but we point out that there exists a physically 
very important constraint on the scalar function $\chi (\bm{x})$ given by
\begin{equation}
 \nabla \times \nabla \chi (\bm{x}) \ = \ 0, \label{Eq:chi_notT}
\end{equation}
which is often forgotten when the gauge ambiguity issue of the
vector potential in the AB effect is discussed. 
As we shall soon argue in more detail, if this condition is not satisfied, 
the part $\nabla \chi (\bm{x})$ of $\bm{A} (\bm{x})$ would generate 
a new magnetic field distribution which necessarily alters the original 
distribution $\bm{B} (\bm{x})$.

At the moment, let us go ahead by assuming that the condition (\ref{Eq:chi_notT}) is 
satisfied anyhow. Then, if it is combined with the easily verified another relation
$\nabla \cdot \bm{A}^{(S)} (\bm{x}) = 0$, Eq.(\ref{Eq:A_S+nabla_chi}) 
just gives the transverse-longitudinal decomposition of the vector 
potential \cite{Adachi1992}\nocite{Stewart2003}-\cite{Li2012}
\begin{equation}
 \bm{A} (\bm{x}) \ = \ \bm{A}_\perp (\bm{x}) \ + \ \bm{A}_\parallel (\bm{x}),
\end{equation}
with the following identification
\begin{equation}
 \bm{A}_\perp (\bm{x}) \ \equiv \ \bm{A}^{(S)} (\bm{x}), \ \ \  
 \bm{A}_\parallel (\bm{x}) \ \equiv \ 
 \nabla \, \chi (\bm{x}).
\end{equation}
In fact, these two components certainly satisfy the transverse condition and 
the longitudinal condition, respectively 
\begin{equation}
 \nabla \cdot \bm{A}_\perp (\bm{x}) \ = \ 0, \ \ \ 
 \nabla \times \bm{A}_\parallel (\bm{x}) \ = \ 0.
\end{equation}
Note that the derivation above indicates that, in our setting of an 
infinitely-long solenoid, the transverse-longitudinal decomposition of the 
vector potential is unique \cite{Adachi1992}\nocite{Stewart2003}-\cite{Li2012}. 
(Note that this is equivalent to saying that the transverse 
part of the vector potential is unique. To avoid misunderstanding, however,
we recall in Appendix \ref{AppendixA} that there is one familiar physical system in which
the transverse-longitudinal decomposition of the vector potential is not 
unique at all.)

Unfortunately, the uniqueness of the transverse-longitudinal decomposition of the
vector potential has been often suspected, probably because of the existence 
of the following gauge transformation \cite{BL1978}
\begin{equation}
 \bm{A}^\prime (\bm{x}) \ = \ \bm{A} (\bm{x}) \ + \ \nabla \chi^{sing} (\bm{x}),
 \label{Eq:A^prime} 
\end{equation}
which is specified by the multi-valued gauge function as follows
\begin{equation}
 \chi^{sing} (\bm{x}) \ = \ - \,\frac{1}{2 \,\pi} \,\Phi \,\phi \ = \ 
 - \,\frac{1}{2 \,\pi} \,\Phi \,\arctan \left( \frac{y}{x} \right), \label{Eq:chi^sing}
\end{equation}
Here $\Phi = \pi \,R^2 \,B$ is the total magnetic flux penetrating the solenoid.
As can be easily verified, the rotation of $\chi^{sing} (\bm{x})$ does not vanish, i.e.
$\nabla \times \nabla \chi^{sing} (\bm{x}) \neq 0$, but it rather satisfies the
{\it transverse condition} $\nabla \cdot \nabla \chi^{sing} (\bm{x}) = 0$.
At first sight, this observation appears to show that the 
transverse-longitudinal decomposition, or equivalently, the identification of 
$\bm{A}^{(S)} (\bm{x})$ as the transverse component is not unique at all, once 
the multi-valued gauge transformation as above is permitted.

 We however recall that our discussion of the relation (\ref{Eq:A_S+nabla_chi})
 based on the Bio-Savart law already indicates that the scalar function
 $\chi (\bm{x})$ in this equation must satisfy the rotation free condition
 $\nabla \times \nabla \chi (\bm{x}) = 0$.
 Otherwise, the term $\nabla \chi (\bm{x})$ in $\bm{A} (\bm{x})$ 
 inevitably alters the magnetic field distribution of the system.
 Let us look into this state of affairs in more concrete manner. 
 As is well-known, the vector potential $\bm{A}^{(S)} (\bm{x})$ obtained from
 the integral (\ref{Eq:A_S_int}) is given by (see page 208 of 
 \cite{Shadowitz1975}, for example)
 \begin{equation}
  \bm{A}^{(S)} (\bm{x}) \ = \ \left\{ \begin{array}{ll}
  \frac{\Phi}{2 \,\pi} \,\frac{\rho}{R^2} \,\bm{e}_\phi \ & \ 
  (\rho < R) \\
  \frac{\Phi}{2 \,\pi} \,\frac{1}{\rho} \,\bm{e}_\phi \ & \ 
  (\rho \geq R) \,\, .
  \end{array} \right. 
 \end{equation}
It is an elementary exercise to get the explicit form of the gauge-transformed
vector potential $\bm{A}^\prime (\bm{x})$ given by (\ref{Eq:A^prime}).
First, for $\rho \geq R$, i.e. in the outer region of the solenoid, we get
\begin{equation}
 \nabla \chi^{sing} (\bm{x}) \ = \ - \,\frac{\Phi}{2 \,\pi} \,\frac{1}{\rho} \,
 \bm{e}_\phi ,
\end{equation}
which in turn gives
\begin{equation}
 \bm{A}^\prime (\bm{x}) \ = \ \frac{\Phi}{2 \,\pi} \,\frac{1}{\rho} \, \bm{e}_\phi
 \ - \ \frac{\Phi}{2 \,\pi} \,\frac{1}{\rho} \, \bm{e}_\phi \ = \ 0 \ \ \ \ \ 
 (\mbox{for} \ \rho \geq R).
\end{equation}
This means that, by the above multi-valued gauge transformation,
the vector potential outside the solenoid can be completely eliminated.
At first sight, this appears to show the expulsion of the AB effect,
as advocated by Bocchieri and Loisinger many years ago \cite{BL1978}. 
However, as was shown later by several researchers \cite{BB1983},\cite{Riess1972},
the AB effect remains 
to exist even after such a multi-valued gauge transformation,
if one properly takes account of the change of the $2 \,\pi$ periodic boundary
condition of the electron wave functions \footnote{See also \cite{Kreizschmar1965}
concerning the general consideration of multi-valued wave functions.} .

Let us next examine what happens with the transformed vector potential inside 
the solenoid. First, in the domain excluding the origin ($\rho = 0$), we find that 
\begin{eqnarray}
 \nabla \chi^{sing} (\bm{x}) &=& - \,\frac{\Phi}{2 \,\pi} \,\frac{1}{\rho} \,
 \bm{e}_\phi \ \ \ \ \ (\rho \neq 0), \\
 \bm{A}^\prime (\bm{x}) &=& \frac{\Phi}{2 \,\pi} \,\left(
 \frac{\rho}{R^2} \ - \ \frac{1}{\rho} \right) \,\bm{e}_\phi \ \ \ \ (\rho \neq 0), \\
 \nabla \times \bm{A}^\prime (\bm{x}) &=& \frac{\Phi}{\pi \,R^2} \,\bm{e}_z
 \ \ \ \ \ (\rho \neq 0).
\end{eqnarray}
The form of $\nabla \chi^{sing} (\bm{x})$ above indicates that
$\nabla \times \nabla \chi^{sing} (\bm{x})$ has a singularity at the origin.
To confirm it, let us consider a circle $C_\varepsilon$ around the origin 
with infinitesimally small radius 
$\varepsilon \, ( \, \rightarrow 0^+ )$. 
The area surrounded by $C_\varepsilon$ is denoted as $S_\varepsilon$. 
If we evaluate the surface integral of 
$\nabla \times \nabla \chi^{sing} (\bm{x})$ over 
$S_\varepsilon$ with the use of the Stokes theorem, we obtain
\begin{eqnarray}
 \int_{S_\varepsilon} \,\left( \nabla \times \nabla \chi^{sing} (\bm{x}) \right)
 \cdot d \bm{S} &=&
 \oint_{C_\varepsilon} \nabla \chi^{sing} (\bm{x}) \cdot d \bm{x} \nonumber \\
 &=& - \,\frac{\Phi}{2 \,\pi} \,\int_0^{2 \,\pi} \frac{1}{\varepsilon} \,\varepsilon \,
 d \phi \ = \ - \,\Phi .
\end{eqnarray}
This indicates that, in the vicinity of the origin, the following relation holds
\begin{equation}
 \nabla \times \nabla \chi^{sing} (\bm{x}) \ = \ 
 - \,\frac{\Phi}{2 \,\pi} \,\frac{\delta (\rho)}{\rho} \,\bm{e}_z .
\end{equation}
In fact, it can be verified from the following manipulation,
\begin{equation}
 \int_{S_\varepsilon} \,\left( \nabla \times \nabla \chi^{sing} (\bm{x}) \right)
 \cdot d \bm{S} \ = \ - \ \frac{\Phi}{2 \,\pi} \,\int_0^{2 \,\pi} \, d \phi \,
 \int_0^\varepsilon \,\frac{\delta (\rho)}{\rho}
 \,\rho \,d \rho \ = \ - \,\Phi.
\end{equation}
To sum up, inside the solenoid, we find that
\begin{eqnarray}
 \bm{B}^\prime (\bm{x}) \ = \ \nabla \times \bm{A}^\prime (\bm{x})
 &=&  \frac{\Phi}{\pi \,R^2} \,\bm{e}_z \ - \ 
 \frac{\Phi}{2 \,\pi} \,\frac{\delta (\rho)}{\rho} \,\bm{e}_z \nonumber \\
 &\equiv& \ \ \ \bm{B} (\bm{x}) \ \ + \ \ \bm{B}^{string} (\bm{x})  
 \hspace{8mm} (\rho < R) .
\end{eqnarray}
The 1st term of the above equation is nothing but the original uniform magnetic 
field $\bm{B} (\bm{x})$ inside the solenoid. 
On the other hand, the 2nd term shows that the multi-valued gauge
transformation generates a string-like magnetic field directed to the negative
$z$-axis, which is opposite to the direction of the
original uniform magnetic field. In this way, as pointed out before,
we confirm that that the multi-valued gauge transformation specified by
(\ref{Eq:A^prime}) and (\ref{Eq:chi^sing}) generates an extra magnetic
field distribution which is originally absent. 
If we evaluate the total flux of $\bm{B}^\prime (\bm{x})$
penetrating the solenoid (with the radius $R$), we obtain
\begin{eqnarray}
 \int_{S (\rho \leq R)} \,\bm{B}^\prime (\bm{x}) \cdot \bm{n} \,d S &=& 
 \int_{S (\rho \leq R)} \,\frac{\Phi}{\pi \,R^2} \,\bm{e}_z \cdot \bm{e}_z \,d S
 \ - \ \int_{S (\rho \leq R)} \,\frac{\Phi}{2 \,\pi} \,
 \frac{\delta (\rho)}{\rho} \,\bm{e}_z \cdot \bm{e}_z \,d S \hspace{5mm}
 \nonumber \\
 &=& \ \ \ \Phi \ \ - \ \ \Phi \ = \ 0,
\end{eqnarray}
which means that the new net magnetic field penetrating the solenoid becomes 
precisely zero. Undoubtedly, this is the reason why the gauge-transformed 
vector potential $\bm{A}^\prime (\bm{x})$ exactly vanishes outside the 
solenoid.

\vspace{2mm}
Unphysical nature of such a singular gauge
transformation can also be convinced if we evaluate the curl of 
$\bm{B}^\prime (\bm{x}) \equiv
\bm{B} (\bm{x}) + \bm{B}^{string} (\bm{x})$.
We find that
\begin{eqnarray}
 \nabla \times \bm{B} (\bm{x}) &=& B \,\delta (\rho - R) \,\bm{e}_\phi
 \ = \ \bm{J}_{ext} (\bm{x}) \\
 \nabla \times \bm{B}^{string} (\bm{x}) &=& 
 \frac{\Phi}{2 \,\pi} \,\frac{\partial}{\partial \rho} \,
 \left( \frac{\delta (\rho)}{\rho} \right) \,\bm{e}_\phi
 \ \equiv \ \bm{J}_{string} (\bm{x}).
\end{eqnarray}
This means that the new magnetic field $\bm{B}^\prime (\bm{x})$ satisfies the
following equation
\begin{equation}
 \nabla \times \bm{B}^\prime (\bm{x}) \ = \ \bm{J}_{ext} (\bm{x}) \ + \ 
 \bm{J}_{string} (\bm{x}) .
\end{equation}
This is clearly different from the original Maxwell equation for the magnetic field
$\bm{B} (\bm{x})$ given by
\begin{equation}
 \nabla \times \bm{B} (\bm{x}) \ = \ \bm{J}_{ext} (\bm{x}).
\end{equation}
Beyond doubt, all these observations reveal physically unacceptable nature of the 
above multi-valued gauge transformation.
We therefore conclude that, as long as such an unphysical gauge transformation 
is excluded, the transverse-longitudinal decomposition of the vector potential
is unique at least in our setting of an infinitely-long solenoid.

\section{Unveiling the role of vector potential in the Aharonov-Bohm effect}\label{sect3}

Through the discussion in the previous sections, we have verified that,
at least in the setting of an infinitely-long solenoid, the generated vector potential
can uniquely be decomposed into the transverse component and the longitudinal 
component, provided that the possibility of multi-valued gauge transformation 
is excluded.
The point is that the multi-valued gauge transformation may be mathematically 
allowed, but it is physically unacceptable because it alters the magnetic field 
distribution of the system or even the form of the basic Maxwell equation.

Most importantly, the transverse component of the vector potential, i.e.
$\bm{A}_\perp (\bm{x}) = \bm{A}^{(S)} (\bm{x})$ is unique, gauge-invariant and 
it cannot be eliminated by any regular gauge transformations which leave
the magnetic field distribution intact. 
It strongly indicates that this transverse component
of the vector potential is not just a convenient mathematical tool but
it contains some definite physical entity. Nevertheless, it is also true that the entire
vector potential still contains the longitudinal part, which cannot be free
from gauge ambiguity. How one should confront this puzzling
situation has already been argued by several 
researchers \cite{Adachi1992}\nocite{Stewart2003}-\cite{Li2012}.
Unfortunately, in any of these previous investigations, the uniqueness argument
of the transverse-longitudinal decomposition has not been completed at the 
satisfactory level as discussed in the present paper. 
Probably, this is the reason why such past analyses could not
completely dispel the misbelief that the vector potential is just a 
mathematical tool with little physical substance.
Now, we are ready to make more definitive statement on this long-standing 
frustrating situation.

\begin{figure}
\centering
\includegraphics[width=5.5cm]{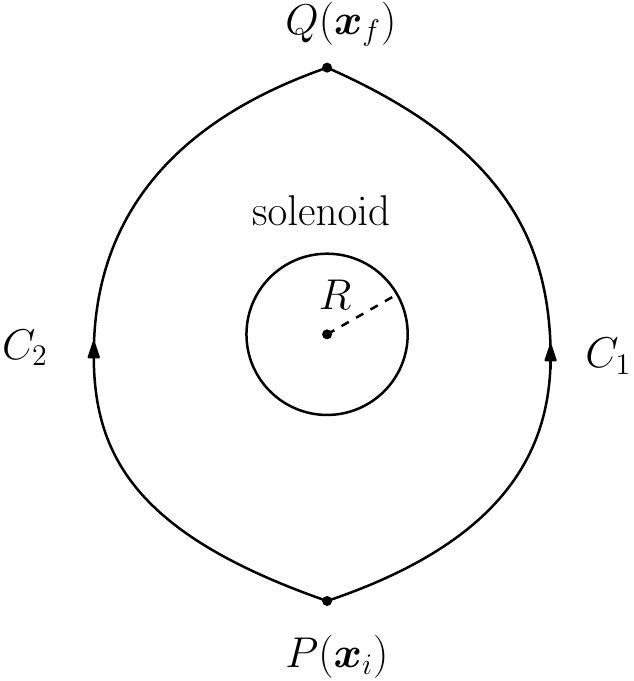}
\caption{Schematic picture showing two paths connecting the initial point $P (\bm{x}_i)$
where the electron beam is ejected and the final point $Q (\bm{x}_f)$ on the screen.}
\label{Two_paths}
\end{figure}

First, let us consider the most fundamental AB-phase shift 
measured through the interference of the two electron beams.
(See the schematic picture illustrated in Fig.{\ref{Two_paths}}.)
According to the standard analysis, the phase change of the electron wave function
along the path $C_1$ is given by
\begin{equation}
 \Delta \phi_{AB} (C_1) \ = \ e \,\int_{C_1} \,\bm{A} (\bm{x}) \cdot d \bm{x},
\end{equation}
while the phase change along the path $C_2$ is given by
\begin{equation}
 \Delta \phi_{AB} (C_2) \ = \ e \,\int_{C_2} \,\bm{A} (\bm{x}) \cdot d \bm{x}.
\end{equation}

Since the observable AB phase shift corresponds to the difference between the
above two phase shifts, it is eventually given by the following expression,
which is proportional to the closed line integral of the vector potential
$\bm{A} (\bm{x})$ represented as
\begin{equation}
 \phi_{AB} \ = \ e \,\int_{C_1} \,\bm{A} (\bm{x}) \cdot d \bm{x} \ - \ 
 e \,\int_{C_2} \,\bm{A} (\bm{x}) \cdot d \bm{x} 
 \ = \ e \,\oint_{C_1 - C_2} \,\bm{A} (\bm{x}) \cdot d \bm{x} .
\end{equation}
By now, we know that the vector potential is generally given as
$\bm{A} (\bm{x}) = \bm{A}^{(S)} (\bm{x}) + \nabla \chi (\bm{x})$.
The above closed line integral of the vector potential is then given by
\begin{equation}
 \oint_{C_1 - C_2} \bm{A} (\bm{x}) \cdot d \bm{x} \ = \ 
 \oint_{C_1 - C_2} \bm{A}^{(S)} (\bm{x}) \cdot d \bm{x} \ + \ 
 \oint_{C_1 - C_2} \nabla \chi (\bm{x}) \cdot d \bm{x}. 
\end{equation}
Here we have
\begin{equation}
 \oint_{C_1 - C_2} \bm{A}^{(S)} (\bm{x}) \cdot d \bm{x} \ = \ 
 \int_S \bm{B} (\bm{x}) \cdot d \bm{S} \ = \ \Phi,
\end{equation}
and
\begin{equation}
 \oint_{C_1 - C_2} \nabla \chi (\bm{x}) \cdot d \bm{x} \ = \ 
 \int_S \left( \nabla \times \nabla \chi (\bm{x}) \right) \cdot d \bm{S}
 \ = \ 0,
\end{equation}
since the gauge function $\chi (\bm{x})$ is demanded to satisfy the
constraint $\nabla \times \nabla \chi (\bm{x}) = 0$.
This means that the transverse component $\bm{A}^{(S)} (\bm{x})$ solely
explains the Aharonov-Bohm phase shift and the gauge-dependent
longitudinal component never contributes to it.

Importantly, however, if we consider the phase change corresponding to a
non-closed path connecting the two spatial points $\bm{x}_i$ and $\bm{x}_f$,
it is given by
\begin{equation}
 \Delta \phi_{AB} \ = \ e \,\int_{\bm{x}_i}^{\bm{x}_f} \bm{A} (\bm{x}) \cdot d \bm{x} .
\end{equation}
This quantity is also divided into two pieces as
\begin{eqnarray}
 \Delta \phi_{AB} &=&  e \,\int_{\bm{x}_i}^{\bm{x}_f} \,\left( \bm{A}^{(S)} (\bm{x})
 \ + \ \nabla \chi (\bm{x}) \right) \cdot d \bm{x} \nonumber \\
 &=& e \,\int_{\bm{x}_i}^{\bm{x}_f} \bm{A}^{(S)} (\bm{x}) \cdot d \bm{x} 
 \ + \ e \,\left( \chi (\bm{x}_f) \ - \ \chi (\bm{x}_i) \right) .
\end{eqnarray}
Although the 1st term of the above equation is gauge invariant, the 2nd term
is not, because of the arbitrariness of the gauge function $\chi (\bm{x})$ 
contained in the longitudinal part. 
This means that such a {\it partial} AB-phase shift is a gauge-dependent quantity.
Hence, as long as we believes the widely-accepted gauge principle,
we must conclude that it would not correspond to any measurable quantity.
The statement above may sound self-evident to some researchers.
We however recall that this conclusion contradicts the recent claims in several authors
that such {\it partial} Aharonov-Bohm phase shift can in principle be observed.
In the next section, we briefly introduce and comment on these challenging
claims.

\section{On some attempts to explain the AB effect 
without using the gauge-variant electromagnetic potential}\label{sect4}

It is widely accepted that the vital importance of the vector potential
in quantum mechanics was established in the paper by
Aharonov and Bohm on the quantum phenomenon
associated with their names \cite{AB1959}. However, it appears that, because 
of the {\it gauge-dependent} nature of the vector potential, Aharonov himself
is not completely satisfied with the vector potential interpretation
of the AB effect. This motivates him and his collaborators to look
for explanations which do not use the gauge-dependent vector 
potential \cite{AV2000},\cite{ACR2016}.
Along with this line of explorations, Vaidman argued that when the source
of the electromagnetic potential is treated in the framework of quantum 
theory, the AB effect can be explained without the notion of vector 
potential \cite{Vaidman2012}.
To be more concrete, he considered the following setup.
The solenoid consists of two cylinders and the opposite charges
$Q$ and $- \,Q$ spread on their surfaces. The cylinders rotate in
opposite directions with a certain surface velocity.
The electron is supposed to encircle the solenoid with some
velocity in a superposition of being in the left and in the right sides 
of the circular trajectory.
When the electron enters one arm of the circle, it changes the
magnetic flux through a cross section of the solenoid and then
induces an electromagnetic force acting on the solenoid.
Vaidman explicitly calculated the shift of the wave packet of each
cylinder during the motion of the electron. Combining this results with
the information on the relevant wavelength of the de Broglie wave 
of each cylinder, he eventually showed that this analysis precisely
reproduces the familiar AB phase shift. 

Although Vaidman emphasized the quantum mechanical nature of
his analysis, there appears to be close connections between
his analysis and Boyer's semiclassical 
analysis of the AB effect \cite{Boyer2002},
especially as to the basic interaction dynamics of the solenoid and electron.
Boyer considered a solenoid as a stack of electric current loops.
He calculated Lorentz force due to the electron acting on charge
carriers flowing in each current loop.
This Lorentz force was shown to generate the change in velocity
and the electron paths. By calculating the difference in path length
for electron paths passing on either side of the solenoid, Boyer
demonstrated that the resultant path length difference leads to a
semiclassical phase shift which reproduces the known AB-phase shift.

Anyhow, a common ingredient in the explanations of Vaidman and of Boyer is
the presence of force acting on the solenoid induced by the motion of 
the electron. However, we recall that such an explanation of 
the AB-phase shift due to force has been believed to be incompatible with 
the dispersionless nature of the AB effect, which means that the magnitude
of the AB-phase shift is independent of the electron 
velocity \cite{Zellinger1986},\cite{Peshkin1999}. 
Unfortunately, no decisive experiment to verify the dispersionless nature 
of the AB effect has been carried out for a long time. 
Some years ago, however, Caprez, Barwick, and Batelaan carried
out a crucial time delay measurement of the electron beam and
verified that no time delay was observed, thereby concluding that all force
explanations of the AB-phase shift are ruled out \cite{CBB2007}.

Also worthy of mention is the existence of still another explanation
of the AB-phase shift. 
The basic postulate of this approach is that the AB-phase
shift is proportional to the change of the interaction energy between the
charged particle and the solenoid along the path of the moving charge.
The idea can be traced back to Boyer's old work \cite{Boyer1971}, which 
is based on the framework of classical electrodynamics . 
He assumed that the AB-phase
shift for a given path of the moving charge is proportional to the change
of the interaction energy between the magnetic field $\bm{B}^s$
generated by the current of an infinitely-long solenoid and the magnetic
field $\bm{B}^\prime$ generated by a moving charge with a constant
velocity $\bm{v}$ as
\begin{equation}
 \Delta \phi_{AB} \ \propto \ \Delta \epsilon \,({\rm Boyer}) \ = \ \int \,
 \bm{B}^s (\bm{x}^\prime) \cdot 
 \bm{B}^\prime (\bm{x}^\prime, t) \,\, d^3 x^\prime ,
\end{equation}
where $\bm{B}^s (\bm{x})$ the magnetic field generated by the surface current
$\bm{j}_{ext} (\bm{x})$ of the solenoid according to the Maxwell equation,
\begin{equation}
 \nabla \times \bm{B}^s (\bm{x}) \ = \ 
 \bm{j}_{ext} (\bm{x}) .
\end{equation}
After transforming the above expression by making full use of the knowledge 
of classical electrodynamics, Boyer arrived at a remarkable relation
\begin{equation}
 \Delta \varepsilon \,({\rm Boyer}) \ = \ e \,\bm{v} \cdot 
 \bm{A}^{(S)} (\bm{x}) ,
\end{equation}
with
\begin{equation}
 \bm{A}^{(S)} (\bm{x}) \ = \ \frac{1}{4 \,\pi} \,\int \, 
 \frac{\bm{j}_{ext} (\bm{x}^\prime)}{\vert \bm{x} - \bm{x}^\prime \vert} \,\,
 d^3 x^\prime.
\end{equation}
As emphasized by Boyer, the above $\Delta \varepsilon \,({\rm Boyer})$ is
free from the gauge choice. This is because, in the above expression, 
the quantity $\bm{A}^{(S)} (\bm{x})$ is uniquely determined by the  surface 
current $\bm{j}_{ext} (\bm{x})$ of the solenoid, which is gauge invariant.
This claim sounds reasonable, because the interaction energy is likely
to be a gauge-invariant quantity.

Motivated by the work of Boyer, several researchers investigated the interaction 
energy between the solenoid and a moving charge within the framework of 
quantum electrodynamics \cite{Santos-Gonzalo1999}\nocite{Kang2017}
\nocite{Marletto-Vedral2020}-\cite{Saldanha2021}.
(Also noteworthy is a related but slightly different approach
discussed in \cite{KMY2024}.) 
They evaluated the interaction energy between the solenoid current and the
charged particle mediated by the exchange of a virtual photon within the
framework of the quantum electrodynamics, thereby arriving at the following
answer,
\begin{equation}
 \Delta \varepsilon \,({\rm virtual \ photon \ exchange}) \ = \  
 - \,e \,\bm{v} \cdot \bm{A}^{(S)} (\bm{x}) ,
\end{equation}
where $\bm{A}^{(S)} (\bm{x})$ is the same quantity as appearing in the
corresponding interaction energy obtained by Boyer \cite{Boyer1971}.

Important messages from the authors of the above investigations are
as follows.
The change of interaction energy between the solenoid and the charged
particle along the path of the moving charge is a gauge-invariant quantity.
Therefore, if one accept the above-mentioned postulate that the phase change
of the electron wave function is proportional to the change of interaction energy 
along the path of the moving charge, the AB-phase shift for a {\it non-closed path} 
is also a gauge-invariant quantity. This appears to indicate that it can in principle 
be observed. In fact, based on this belief,
several authors proposed some concrete measurements
for extracting the {\it partial} AB-phase shift corresponding to a non-closed 
path \cite{Kang2017}\nocite{Marletto-Vedral2020}-\cite{Saldanha2021}.

This claim was however criticized in a recent paper by 
ourselves \cite{Waka2024}.
It was pointed out that, very strangely, the expressions of the 
interaction energy of Boyer and that due to the virtual-photon exchange are 
just identical with opposite signs\footnote{
This remarkable fact was never noticed before, since the above
researchers paid attention only to the absolute magnitude of the
predicted AB-phase shift. }.
It was further shown that, within the framework of a self-contained quantum
mechanical treatment of the combined system of a solenoid, a
charged particle, and the quantized electromagnetic field, the interaction
energy of Boyer and that due to virtual-photon exchange precisely cancel
to each other. (Since this demonstration requires fairly careful 
preparation, interested readers are recommended to read the original paper 
\cite{Waka2024} .) 
The analysis there rather shows that the origin of
the AB-phase shift can be traced back to other part of the
self-contained treatment above, which  is after all nothing different from the 
standard mechanism as explained in sect.\ref{sect3} of the present paper. 
This means that the AB-phase shift corresponding to a non-closed path is 
not a gauge-invariant quantity so that its observation is most likely to 
contradict the celebrated gauge principle.  
In any case, it seems to us that any attempts, which aims at explaining
the AB-phase shift without using the notion of the vector potential,
has not been entirely successful up to the present. 
At this point in time, the vector potential interpretation seems to be the 
simplest and most reasonable physical explanation of the AB effect.

\section{Summary and conclusion}\label{sect5} 

The vector potential interpretation of the AB effect is not
universally accepted because of the gauge-variant nature of the
vector potential. 
Even now, not a few researchers seem to believe that the vector potential is 
just a convenient tool for obtaining the electromagnetic field,
and they are searching an explanation of the AB effect without
using the vector potential concept. In the present paper, we tried to convince that 
the vector potential is not just a convenient mathematical
tool with little physical substance. The argument proceeds as follows.
Employing the simplest setting of the system, i.e. an infinitely-long solenoid,
we have shown the following facts, 

\begin{itemize}
\item The vector potential generated by the infinitely-long solenoid is given as
a sum of the transverse part and the longitudinal part.

\item The above decomposition is unique as far as the multi-valued
gauge transformation is excluded. In particular, the transverse part of the vector
potential is uniquely determined by the surface electric current distribution of
the solenoid.  

\item The multi-valued (and singular) gauge transformation is not allowed from
the physical point of view, because it inevitably generates a new or
extra magnetic field distribution which is originally absent in the system.

\item The transverse part of the vector potential solely explains the standard
Aharonov-Bohm effect corresponding to a closed path of the electron's trajectory.

\item Nevertheless, one should not forget the fact that the vector potential
still contains the longitudinal part which has inherent gauge arbitrariness. 
It seems to us that this gauge arbitrariness forbids the observation of the partial 
AB-phase shift corresponding to a non-closed path, which was recently claimed 
to be possible by several researchers.
\end{itemize} 

\vspace{2mm}
To sum up, we conclude that the vector potential contains in it a piece which
is unique, gauge-invariant and cannot be eliminated by any regular gauge transformations.
This part of the vector potential solely explains the standard Aharonov-Bohm
effect. However, still remaining ambiguity of the longitudinal part of the vector potential 
is thought to forbid the observability of the {\it partial} Aharonov-Bohm phase shift 
corresponding to a non-closed path, because it is gauge-dependent
and its observation contradicts the celebrated gauge principle.
Conversely speaking, if the AB-phase shift corresponding to a non-closed path
were observed, it would give us first counterexample to the validity of the gauge 
principle. Undoubtedly, this last statement deeply concerns the authenticity
of the vector potential interpretation of the Aharonov-Bohm effect 
considerably refined in the present paper.

%






\appendix
\section[\appendixname~\thesection]{One familiar physical system in which the 
transverse-longitudinal decomposition of the vector potential is not unique}\label{AppendixA}


\vspace{1mm}
In sect.\ref{sect2}, we have shown that the magnetic vector 
potential generated by
an infinitely-long solenoid is uniquely decomposed into the transverse and longitudinal
components, once we exclude physically unacceptable multi-valued or singular gauge 
transformation.
Naturally, whether the transverse-longitudinal decomposition of the vector potential is
unique or not depends on what physical system we are considering.
One interesting example is provided by the familiar Landau problem, which 
handles the quantum mechanical motion of an electron in an infinitely spreading
uniform magnetic field.
As is well-known, there are three typical choices of gauge potential (configuration)
which reproduce the uniform magnetic field.
They are the symmetric gauge potential $\bm{A}^{(S)} (\bm{x})$, the 1st Landau gauge
potential $\bm{A}^{(L_1)} (\bm{x})$, and the 2nd Landau gauge potential 
$\bm{A}^{(L_2)} (\bm{x})$ respectively given as
\begin{eqnarray}
 \bm{A}^{(S)} (\bm{x}) &=& \frac{1}{2} \,\left( - \,B \,y \,\bm{e}_x + B \,x \,\bm{e}_y \right)
 \ = \ \frac{1}{2} \,B \, r \, \bm{e}_\phi, \\
 \bm{A}^{(L_1)} (\bm{x}) &=& - \,B \,y \,\bm{e}_x, \\
 \bm{A}^{(L_2)} (\bm{x}) &=& + \,B \,x \,\bm{e}_y .
\end{eqnarray}
These potentials are related through the following gauge transformations
\begin{eqnarray}
 \bm{A}^{(S)} (\bm{x}) &=& \bm{A}^{(L_1)} (\bm{x}) \ + \ \nabla \chi_1 (\bm{x}) 
 \ \ \ \mbox{with} \ \ \ \chi_1 (\bm{x}) = + \frac{1}{2} \,B \,x \,y, \\
 \bm{A}^{(S)} (\bm{x}) &=& \bm{A}^{(L_2)} (\bm{x}) \ + \ \nabla \chi_2 (\bm{x}) 
 \ \ \ \mbox{with} \ \ \ \chi_2 (\bm{x}) = - \frac{1}{2} \,B \,x \,y .
\end{eqnarray}
We start with the fact that any gauge potential reproducing the uniform magnetic
field can anyhow be expressed in the form
\begin{equation}
 \bm{A} (\bm{x}) \ = \ \bm{A}^{(S)} (\bm{x}) \ + \ \nabla \chi (\bm{x}),
 \label{Eq:A_AS}
\end{equation}
where $\chi (\bm{x})$ is an arbitrary scalar function subject to the constraint
$\nabla \times \nabla \chi (\bm{x}) = 0$.

With the identification $\bm{A}_\perp (\bm{x}) = \bm{A}^{(S)} (\bm{x})$ and 
$\bm{A}_\parallel (\bm{x}) = \nabla \chi (\bm{x})$, (\ref{Eq:A_AS}) certainly gives a
transverse-longitudinal decomposition of the vector potential as
\begin{equation}
 \bm{A} (\bm{x}) \ = \ \bm{A}^{(S)} (\bm{x}) \ + \ \nabla \chi (\bm{x})
 \ \equiv \ \bm{A}_\perp (\bm{x}) \ + \ \bm{A}_\parallel (\bm{x}) .
\end{equation}
However, the vector potential $\bm{A} (\bm{x})$ can also be expressed in either
of the following forms
\begin{eqnarray}
 \bm{A} (\bm{x}) &=& \bm{A}^{(S)} (\bm{x}) \ + \ \nabla \chi (\bm{x}) , \\
 &=& \bm{A}^{(L_1)} (\bm{x}) \ + \ \nabla \chi^\prime_1 (\bm{x}) \ \ \ \mbox{with} \ \ \ 
 \chi^\prime_1 \ \equiv \ \chi \ + \ \chi_1 , \\
 &=& \bm{A}^{(L_2)} (\bm{x}) \ + \ \nabla \chi^\prime_2 (\bm{x}) \ \ \ \mbox{with} \ \ \ 
 \chi^\prime_2 \ = \ \chi \ + \ \chi_2 . 
\end{eqnarray}
Any of these three give transverse-longitudinal decompositions, since it holds that
\begin{eqnarray}
 \nabla \cdot \bm{A}^{(S)} (\bm{x}) &=& \nabla \cdot \bm{A}^{(L_1)} (\bm{x}) \ = \ 
 \nabla \cdot \bm{A}^{(L_2)} (\bm{x}) \ = \ 0, \\
 \nabla \times \nabla \chi (\bm{x}) &=& \nabla \times \nabla \chi^\prime_1 (\bm{x})
 \ = \ \nabla \times \nabla \chi^\prime_2 (\bm{x}) \ = \ 0.
\end{eqnarray}
Undoubtedly, the transverse-longitudinal decomposition of the vector potential is not
unique in the setting of the Landau system.

Also worthy of mention is the existence of the multi-valued gauge transformation
in the Landau system \cite{WKZ2018}.
Let us consider the gauge potential $\bm{A}^{(BB)} (\bm{x})$ obtained from the
symmetric gauge potential $\bm{A}^{(S)} (\bm{x})$ by the following
multi-valued gauge transformation
\begin{equation}
 \bm{A}^{(BB)} (\bm{x}) \ = \ \bm{A}^{(S)} (\bm{x}) \ + \ \nabla \tilde{\chi} (\bm{x}),
\end{equation}
with
\begin{equation}
 \tilde{\chi} (\bm{x}) \ = \ - \,\frac{1}{2} \,B \,r^2 \,\phi .
\end{equation}
An explicit calculation gives
\begin{equation}
 \bm{A}^{(BB)} (\bm{x}) \ = \ - \,B \,r \,\phi \,\bm{e}_r,
\end{equation}
which was called the vector potential in the Bawin-Burnel gauge \cite{BB1983} 
in \cite{WKZ2018}.
Different from the infinitely-long solenoid problem, the above multi-valued
gauge transformation does not generate an extra magnetic field distribution,
because the above function $\nabla \tilde{\chi} (\bm{x})$ satisfies 
rotation free condition
\begin{equation}
 \nabla \times \nabla \tilde{\chi} (\bm{x}) \ = \ 0.
\end{equation}
We emphasize that this situation is significantly different from the case of the 
multi-valued gauge transformation in the infinitely-long solenoid system, 
which inevitably generates a new or extra magnetic field distribution of string type.

\vspace{2mm}
Finally, for reference, we remind of the fact that non-uniqueness
of the transverse-longitudinal decomposition of the vector potential in the Landau 
problem is related to a special nature of the Landau system, in which the uniform
and constant magnetic field is spreading over whole the $x$-$y$ plane. 
It is obvious that it breaks the validity conditions of the famous Helmholtz
theorem, which insures the uniqueness of the transverse-longitudinal
decomposition of a vector field (see Appendix B of the book \cite{Griffiths1999},
for example).

\vspace{3mm}





\end{document}